\documentstyle[aps,prb,multicol,eqsecnum]{revtex}
\input{psfig}
\begin{document}
\newcommand{\vecr}{{\bf r}}
\newcommand{\vecy}{{\bf y}}
\newcommand{\veck}{{\bf k}}
\newcommand{\vecv}{{\bf v}}
\newcommand{\vecE}{{\bf E}}
\newcommand{\vecj}{{\bf j}}
\newcommand{\kperp}{{{\bf k}_\perp}}
\newcommand{\sinch}{{\rm sinch}}
\newcommand{\cth}{{\rm cth}}
\newcommand{\bartau}{{{\bar \tau}_T}}
\renewcommand{\Re}{{\rm Re}}
\newcommand{\spc}{{\,\,\,\,\,\,\,\,}}
\newcommand{\bea}{\begin{eqnarray}}
\newcommand{\eea}{\end{eqnarray}}
\renewcommand{\[}{\begin{equation}} 
\renewcommand{\]}{\end{equation}}
\newcommand{\bef}{\begin{figure}} 
\newcommand{\ef}{\end{figure}}
\newcommand{\ie}{{\it i.e.}}
\newcommand{\eg}{{\it e.g.}}
\newcommand{\llabel}[1]{\label{#1}}
\newcommand{\eq}[1]{Eq.~(\ref{#1})} 
\newcommand{\fig}[1]{Fig.~\ref{#1}} 

\title{Noise properties and ac conductance of
mesoscopic diffusive conductors with screening}
\author{Y. Naveh, D. V. Averin, and K. K. Likharev}
\address{Department of Physics, State University 
of New York, Stony Brook, NY 11794-3800}
\date{\today}
\maketitle

\begin{abstract}
A theory of non-equilibrium (``shot'') noise and high frequency conductance in
diffusive mesoscopic conductors with screening is presented. Detailed
results are obtained for two simple geometries, for both large and
short electron-electron scattering length $l_{ee}$, at frequencies of
the order of the inverse Thouless time $1/\tau_T$. The  conductance and
the noise are found to exhibit significant 
frequency dependence. For $L \ll l_{ee}$, the high-frequency
($\omega\tau_T \gg 1$) shot
noise spectral density $S_I(\omega)$ approaches a finite value between
$2eI/3$ and $2eI$, depending on the screening properties of the
system, with temperature corrections to $S_I(\omega)$ being {\it
linear} in $T$. However, when $L \gg l_{ee}$, $S_I(\omega)$ grows as
$\omega^{1/4}$ (at $T=0$), is not upper-bound by $2eI$, and has a
temperature-dependent component {\it quadratic} in
$T$. As a result, measurements of $S_I(\omega, T)$ can be utilized as
a probe of the strength of electron-electron scattering.
\end{abstract}

\begin{multicols}{2}
\section{introduction}

Significant attention has recently been focused on the dynamic
electronic properties of mesoscopic systems. These properties include
the ac conductance which gives the mean current response to applied ac
voltage, and the noise, {\it i.e.}, the deviations
of the current from its average value.  Due to the
fluctuation-dissipation theorem, 
equilibrium Johnson-Nyquist noise, as measured in the external leads,
does not convey any additional information to that obtained from the
ac conductance. This is not the case for non-equilibrium ('shot')
noise. Here, the current fluctuations are dependent upon 
the non-equilibrium distribution function, as well as on
electron-electron correlations. Moreover, 
the shot noise may be interpreted as an indication that the transport
mechanism through the structure involves discrete transfer of charge,
as opposed to the continuous charge transfer that takes place in
macroscopic conductors.\cite{Averin 91}

Earlier, shot noise in diffusive conductors was calculated in the
zero-frequency limit,\cite{Beenakker 92,Nagaev 92,Nazarov
94,de_Jong 96} with the conclusion that
the low frequency spectral density equals 1/3 of
the classical Schottky value $2eI$, where $I$ is the average (dc) current
in the system. This result was obtained in two very different
theoretical frameworks, namely in a quantum mechanical transmission
approach\cite{Beenakker 92} which is generally based on the quantum coherence
of the  different electron states, and in a semi-classical
approach,\cite{Nagaev 92} in which quantum coherence is neglected, and
the only effects on the noise are due to the single-particle
non-equilibrium distribution function of the electrons. We believe
that this surprising
agreement between the two theories was adequately explained\cite{de_Jong 94}
by showing that the main ingredient in the transmission
approach, the probability distribution of transmission coefficients, is
not affected by quantum interference in diffusive conductors, thus
establishing the validity of the semi-classical method.\cite{Landauer
96} The 1/3
suppression result is strictly valid only in a non-interacting electron
picture. The electron-electron interaction slightly increases the
zero-frequency noise value, which reaches
$(\sqrt{3}/2) eI$ in the limit when the electrons are locally
thermalized.\cite{Nagaev 95,Kozub 96}

One of the most important characteristics of classical shot noise is
that it is white up to very high frequencies. In ballistic structures
the noise is reduced only at frequencies of the order of the inverse
time of flight of the electron across the device.\cite{van_der_Ziel
54} In diffusive conductors, there are at least three {\it a priori}
candidates to the analog of this time constant: the elastic scattering
time $\tau$, the (much larger) ``Thouless'' time $\tau_T$ of electron
diffusion through the sample, and the Maxwell relaxation time
$(4\pi\sigma)^{-1}$.  In addition, quantum effects also may be
manifested by a frequency-dependent noise spectrum, increasing the
noise at $\omega >~ eV/\hbar$, where $V$ is the applied
voltage.\cite{Altshuler 94} Recently we have shown\cite{Naveh 97} that
even 
at high 
voltage, $eV \gg \hbar/\tau_T$, where quantum effects are
negligible, the shot noise in diffusive structures may exhibit
considerable 
frequency dependence at frequencies as low as
$1/\tau_T$.\cite{Nagaev 97}  In
that work, however, only the zero-temperature case was considered and
the ac conductance and effects of electron-electron
interaction were not explored.

The issue of high-frequency noise cannot be separated from that of the
ac conductance at the same frequency.
Previous works studied the ac conductance in diffusive structures
with ring\cite{Ymri 86,Gefen 91,Reulet 94} or
linear\cite{Buttiker 93,Buttiker 96a,Buttiker 
97} 
geometries. In the first case, non-interacting electrons were considered,
and 
the frequency dependence of the ac 
conductance was found to be similar to the Drude
dependence,\cite{Gefen 91} \ie, appreciable only at $\omega \sim
1/\tau$. However, usual conductors (with electrodes) cannot be
considered separately from their electrodynamic environment.  For such
conductors general expressions for  the 
conductivity 
were obtained to linear order in the frequency.\cite{Buttiker
93,Buttiker 96a,Buttiker 97} Under the assumption of
absolute local electro-neutrality, it was found that this linear
correction (the ``emittance'') vanishes and the conductance is
again independent of the frequency up to $\omega \sim
1/\tau$.\cite{Buttiker 97} Here we are interested in the case where
the conductor's length $L$ or its thickness $t$ are comparable to the
screening length 
$\lambda$, so local 
charge neutrality may no longer be retained. While we confirm the previous
results\cite{Buttiker 97} for $L,t \gg \lambda$, we show that
observable deviations from them may appear already at $L/\lambda \sim 10$
or $t/\lambda \sim 10$.

A vast amount of research has been dedicated to the effects of weak
localization on the dc conductance of mesoscopic diffusive
conductors.\cite{Altshuler 82} Its effect on the ac
conductance\cite{Vitkalov 96} and the high temperature
noise ($T > eV$)\cite{von_Oppen 97} lead to corrections to these
quantities which 
are of the order of the quantum unit of conductance $e^2/h$ (times some
characteristic energy in the case of noise). These
corrections will be neglected in this work, since we will consider the
case 
of conductance much higher than $e^2/h$, so that deviations of the
high-frequency conductance and the noise from their zero-frequency values are
much larger than
the weak localization corrections.

In the present work the noise properties and the ac conductance of
diffusive conductors much shorter than the electron-phonon mean free
path are calculated at frequencies
comparable $1/\tau_T$ or $4\pi \sigma$, with account of screening.
Throughout the work we assume 
that the electrons form a degenerate gas with Fermi wavelength
$\lambda_F$ much
smaller than the elastic mean free path $l$, while $l\ll L,
\lambda$. This allows us to use the 
Boltzmann-Langevin approach introduced by Kogan and Shulman\cite{Kogan
69} (see also Ref.~[24]), and study both the ac conductance
and the noise in a unified way.
In section II we analyze the Boltzmann-Langevin equation in a
non-uniform structure, obtain the boundary conditions for the distribution
function at the conductor-electrode interface, and derive a
``drift-diffusion-Langevin'' equation for the current. In section III
we apply this equation to two specific models of diffusive conductors
(a ``sandwich'' and a conductor over a ground plane). In section IV the
kernels, which describe the response of the system to
external voltage, and to the random Langevin sources, are found. Using
these response functions we calculate the conductance and thermal
noise (section V), and the non-equilibrium shot noise (section
VI). Section VII presents discussion of the results and conclusions.

\section{Boltzmann-Langevin-Poisson theory}

In order to describe both the conductor and electrodes, we need
self-consistent equations for the current in
a system which may be substantially non-uniform on a scale $\Delta
r \gg l$. In the diffusion
approximation the electron distribution 
function can be written as
\[
\label{diffapprx} f = f(\varepsilon, \cos\theta, {\bf r}, t) =
f_s(\varepsilon, {\bf r}, t) + f_a(\varepsilon, {\bf r}, t) \cos \theta, 
\]
where $|f_a| \sim l/L \ll 1$ and $\varepsilon$ is the total electron energy, 
\[
\label{totalepsilon} \varepsilon = \varepsilon_\veck + \varepsilon_c({\bf r}) - e\Phi({\bf  
r},t) \equiv \varepsilon_\veck + U({\bf r},t).
\]
Here $\varepsilon_\veck$ is the kinetic energy of an
electron with momentum ${\bf k}$ while $\varepsilon_c(\vecr)$ is the
equilibrium local 
conduction band 
edge, which includes possible band-bending due to mismatch in
the local Fermi energies in the non-uniform conductor,
and hence describes the equilibrium (``built-in'') electric field
${\bf E}_0 = -\nabla 
\varepsilon_c / e$. $\theta$ is the angle between ${\bf k}$ and the 
direction of the current, and $\Phi({\bf r},t)$ is the time-dependent 
electric potential, so that $U(\vecr, t)$ is the total instantaneous
potential energy of the electrons. In the above variables, the velocity of the
particle is both position- 
and time-dependent, $v = v(\varepsilon, {\bf r},t)$, and the Boltzmann
equation within the usual relaxation-time approximation looks like
\bea
\label{compact} \nonumber
\frac{\partial f}{\partial t} + \frac{\partial f}{\partial
\varepsilon} \frac {\partial U}{  
\partial t} - \frac{\partial f}{\partial
\cos \theta} \frac{\partial \cos \theta}{\partial {\bf k}} \cdot
\nabla U & & \\
+ \nabla f \cdot {\bf v} + \frac{f_a \cos \theta}{\tau} & = & 0 
\eea
where $\tau = \tau({\bf r})$ is the local elastic relaxation
time. Note that in the diffusion
approximation the term 
proportional to $\nabla U$ is usually neglected, since it is of the
order of $f_a^2$. However, in our case 
this term may be linear in $f_a$, because $\nabla U$
has the component $\nabla \varepsilon_c$ even in the absence of current.

As usual,\cite{Lifshitz 81-83} we proceed by separating Eq.~(\ref{compact})
into its symmetric and antisymmetric parts [see \eq{diffapprx}] and in
the first order in $f_a$ we get
\begin{mathletters} \llabel{symmasymm}
\begin{eqnarray}
\frac{\partial f_s}{\partial t} + \frac{\partial f_s}{\partial \varepsilon}
\frac {\partial U}{\partial t}  - f_a \frac{\partial \cos
\theta}{\partial \veck} \cdot  \nabla 
U  & + & \nabla f_a \cdot \vecv \cos \theta = 0, \llabel{symmeq} \\
\left( 
\frac{\partial }{\partial t}  + \frac{1}{\tau} \right) f_a \cos
\theta & + & \nabla f_s
\cdot \vecv = 0. \llabel{asymmeq}
\end{eqnarray}
\end{mathletters}

In this work we are interested only in the case of frequencies 
much smaller than $1/\tau$. In this case Eqs.~(\ref{symmasymm}) may be
combined to 
give
\bea	\label{diffeqfs} \nonumber
\frac{\partial f_s}{\partial t} + \frac{\partial f_s}{\partial
\varepsilon} \frac {\partial U}{\partial t} +
\nabla f_s \cdot \frac{\vecv
\tau}{\cos \theta} 
\frac{\partial \cos \theta}{\partial \veck} \cdot \nabla U  & & \\
- \nabla (\nabla f_s \cdot \vecv \tau)
\cdot \vecv & = & 0.
\eea
Integration of this equation over the directions of ${\bf k}$ gives 
\[ \llabel{generaldiffusion}
\frac{\varepsilon_\veck l}{D} \frac{\partial f_s}{\partial t} -
\nabla \cdot \left[ \varepsilon_\veck l \nabla f_s \right] - \frac 32
l \nabla_\perp f_s \cdot \nabla_\perp U = 0
\]
with $D=D(\vecr) = l(\vecr) v_F(\vecr)/3$ and $l = l(\vecr) =
\tau(\vecr) v_F(\vecr)$, $v_F$ 
being the Fermi velocity. $\nabla_\perp$ denotes differentiation in
the plane perpendicular to the current direction
$x$. \eq{generaldiffusion} is a generalization of the regular
diffusion equation for the distribution function in the case
when the potential or the mean free path change substantially in space.

The random nature of the scattering in the conductor may be
described\cite{Kogan 
69,Kogan 96} by a stochastic source term $J^s({\bf r}, {\bf k}, t)$,
with zero average,  added to
the right hand side of Eq.~(\ref{compact}). Its correlation
function was found by Kogan and Shul'man\cite{Kogan 69} 
assuming Poisson statistics of the scattering events, and taking
into account the Fermi correlations of the electrons. For the
case of strong isotropic impurity scattering ($l \ll L$) the result
reads
\bea
\label{kogancorr} \nonumber \left\langle J_a^s\right.&  &\left. ({\bf r},
\varepsilon, t) J_a^s({\bf 
r}^{\prime},  
\varepsilon^{\prime}, t^{\prime}) \right\rangle = \\
& & \frac{6 \delta({\bf
r}-{\bf r}  
^{\prime}) \delta(\varepsilon-\varepsilon^{\prime}) \delta(t-t^{\prime})}{ 
\tau({\bf r}) {\cal N}({\bf r})} \bar{f}_s(\varepsilon, {\bf r}) \left[ 1 - 
\bar{f}_s(\varepsilon, {\bf r}) \right], 
\eea
where $J_a^s$ is the antisymmetric component of $J^s$, ${\cal N}({\bf r})$
is the local density of states at the Fermi level 
(excluding the spin degeneracy), and $\bar f$ the ensemble
average of the distribution function.

Equations for the current can be obtained by
including the source term $J^s$ in 
equations~(\ref{symmasymm}), and then integrating them over
the electrons' momenta. It is convenient at this stage to change
variables in Eqs.~(\ref
{symmasymm}) from $\varepsilon$ to $\varepsilon_\veck$ 
by using Eq.~(\ref{totalepsilon}). Integrating \eq{symmeq}
over $\veck$,
we get the continuity equation 
\[
\label{continuity}\frac \partial {\partial t}\rho ({\bf r},t)+\nabla \cdot 
{\bf j}({\bf r},t)=0 
\]
with ${\bf j}({\bf r},t)$ the current density, and  $\rho ({\bf r},t)$
the excess 
electron density, {\it i.e.}, the total charge density minus its equilibrium
value $\rho _0({\bf r})$ (which includes the possible charge transfer
when 
two materials have been brought into contact). The Langevin term integrates out from this equation, as
expected for particle-conserving scattering processes.

If, before integration over $\veck$, we multiply Eq.~(\ref{asymmeq})
(including the Langevin term) by $\vecv$, it yields 
\bea
\label{totaldriftdiff} \nonumber {\bf j}({\bf r},t) & = & -D({\bf
r})\nabla 
\left[ \rho _0(  
{\bf r})+\rho ({\bf r},t)\right] \\ 
& & +\left[ \sigma ({\bf r})+\delta \sigma ( 
{\bf r},t)\right] \left[ {\bf E}_0({\bf r})+{\bf E}({\bf r},t)\right]
+ {\bf j}^s({\bf r},t) 
\eea
with $\sigma ({\bf r})=e\tau ({\bf r})\rho _0({\bf r})/m({\bf r})$ and $ 
\delta \sigma ({\bf r},t)=e\tau ({\bf r})\rho ({\bf r},t)/m({\bf r})$.
Here $m( 
{\bf r})$ is  the local  effective mass (for simplicity, the parabolic
and isotropic dispersion relation was assumed), and 
\[
\label{intsources}{\bf j}^s({\bf r},t)=e\tau ({\bf r})\sum_{{\bf k}}{\bf v}_{ 
{\bf k}}{J}^s({\bf r},{\bf k},t). 
\]
At equilibrium, and in the absence of external fluctuation sources,
the current ${\bf j}({\bf r},t)$ should vanish. Thus, the built-in
electric field satisfies the equation 
\[	\llabel{equilibriumconstraint}
-D({\bf r})\nabla \rho _0({\bf r})+\sigma ({\bf r}){\bf E}_0({\bf r})=0
\]
which may be interpreted as the constancy of the electro-chemical
potential at $V=0$ and $j^s = 0$.

The terms
proportional to $\delta \sigma ({\bf r},t)$ in Eq.~(\ref{totaldriftdiff})
are negligible if
\[
\label{smallpotential}e\left| \Phi _0({\bf r})+\Phi ({\bf r},t)\right| \ll
\varepsilon_F-\varepsilon_c({\bf r}), 
\]
(where $\varepsilon_F$ is the equilibrium Fermi energy [Fig.~1(a)]), \ie,
if the band bending and the external potential are small compared to
the local Fermi energy.
Under the condition (\ref{smallpotential}), equation (\ref{totaldriftdiff}),
together with the constraint (\ref{equilibriumconstraint}), yield
the 'drift-diffusion-Langevin' equation
\[
\label{generalj}{\bf j}({\bf r},t)=\sigma ({\bf r}){\bf E}({\bf r},t)-D({\bf  
r}){\bf \nabla }\rho ({\bf r},t)+{\bf j}^s({\bf r},t) 
\]
where the variables include both the deterministic and stochastic
parts. 

The correlation function of the current sources ${\bf j}^s$ in any
direction $\alpha$ follows from equations (\ref{kogancorr}) and
(\ref{intsources}): 
\[
\label{intcorr}\left\langle {j}_\alpha^s({\bf r},t){j}_\alpha^s({\bf
r}^{\prime 
},t^{\prime })\right\rangle =\delta ({\bf r}-{\bf r}^{\prime })\delta
(t-t^{\prime }){\cal S}({\bf r}) 
\]
with the correlator 
\[
\label{SS}{\cal S}({\bf r})=\frac 23e^2\tau ({\bf r}){\cal N}({\bf r})v_F^2( 
{\bf r})\int_0^\infty d\varepsilon \,\bar f_s(\varepsilon ,{\bf r})\left[ 1- 
\bar f_s(\varepsilon ,{\bf r})\right] . 
\]

Now let us consider a system in which a homogeneous conductor connects two
homogeneous electrodes with interfaces at $\pm L/2$, with the only
source of inhomogeneity being the band bending due to charge transfer
between the materials [Fig.~1(a)]. Let
the 
interfaces   be normal to the $x$ axis and much sharper than the
screening lengths $\lambda, \lambda_e$ in the conductor and the
electrodes, respectively. We define the interface regions to be the
regions  of width
$2\delta$ around $\pm L/2$, with $l \ll \delta \ll \lambda, \lambda_e,
L$. A major assumption of this work is that the voltage in the
system drops entirely in the bulk of the conductor, \ie, the resistances
of the electrodes and the electrode-conductor interfaces are 
small compared to that of the conductor. This is 
the natural situation 
when the electrodes are of high conductivity and when the interfaces
are smooth on the scale of $\lambda_F$, so no reflections of electrons
occur at the 
interfaces. 

Let us consider, for example,  the interface at $L/2$ and define
$f_s^c = f_s(L/2 - 
\delta)$ and $f_s^e = f_s(L/2 + \delta)$. Since the voltage drop in the
electrode is negligible, $f_s^e$ is just the equilibrium Fermi-Dirac
distribution 
\[
\label{f0} f_s^e = f_0(\varepsilon+eV/2) = \frac 1{1 + \exp \left(
\frac {\varepsilon  +eV/2 -\mu}T \right) }, 
\]
where the chemical potential $\mu$ is defined as the average of the
chemical potentials in the two electrodes. Moreover, the fact that
there is no voltage drop across the interface region implies that
$f_s^c$ is also given by \eq{f0}, $f_s^c = 
f_s^e$. Integrating this equation over the electron's momenta and
using the relation ${\cal
N}({\bf r}) = 1/4\pi e^2 \lambda^2 ({\bf r})$ leads to the first
boundary condition at the electrode-conductor interface,
\[ \label{bcrho}
\lambda^2 \rho^c = \lambda_e^2 \rho^e
\]
with $\rho^c \equiv \rho(L/2 - \delta)$ and $\rho^e \equiv \rho(L/2 +
\delta)$. 

The second boundary condition is the continuity of the current across
the interface,
\[	\llabel{bcI}
j_{\omega x}^c = j_{\omega x}^e
\]
with $j_{\omega x}^c$,
$j_{\omega x}^e$ the Fourier components of the transverse current
density at $L/2 - 
\delta$, $L/2 + \delta$, 
respectively.\cite{note bc}

If complemented with the 
Poisson equation 
\[
\label{poisson} \nabla \cdot {\bf E}({\bf r},t) = \frac{4\pi}{\epsilon({\bf r 
})} \rho({\bf r},t),
\]
where $\epsilon({\bf r})$ is the dielectric constant,
equations (\ref{continuity}) and (\ref{generalj} -- \ref{poisson})
form a closed system which is the
basis for our calculations.

\section{Models}

\subsection{The sandwich model}

We study two analytically solvable
models which differ in their assumed sample geometry, and hence in
electrostatics.\cite{Naveh 97} In our first, ``sandwich'' model, which is schematically
shown in Fig.~1(b), a short conductor 
of length $L\ll t$ is sandwiched between two wide electrodes ($t$ is
the smallest transverse dimension of the conductor). 
Defining the quantities $\Phi_\omega(x)$,
$q_\omega(x)$, $E_\omega(x)$, $I_\omega(x)$ and $I_\omega^s(x)$ as 
integrals over 
the sample's cross-section of 
the temporal Fourier components of $\Phi(\vecr, t)$,
$\rho(\vecr, t)$, $E_x(\vecr, t)$, $J_x(\vecr, t)$, and $j^s_x(\vecr,
t)$, 
respectively, we get from equations (\ref{continuity},
\ref{generalj}, \ref{poisson}) 
\[	\llabel{1Dcontinuity}
-i\omega q_\omega(x) + \frac {d I_\omega(x)} {d x} = 0,
\]
\[	\llabel{1Dgeneralj}
I_\omega(x) = \sigma(x) E_\omega(x) - D(x) \frac{d q(x)} {d x} +
I_\omega^s(x),
\]
\[	\llabel{1Dpoisson}
\frac {d E_\omega(x)} {d x} = \frac {4 \pi} {\varepsilon(x)}
q_\omega(x). 
\]
[Deriving  \eq{1Dpoisson},
we have neglected the transverse derivatives of $\vecE(\vecr, t)$,
since, by Gauss' theorem, 
they are proportional to the circumference of the cross-section of the
sample, while the derivative in the $x$-direction is proportional to
the cross-section area.]
\begin{figure}[tb]
\centerline{\hspace{-31pt} \psfig{figure=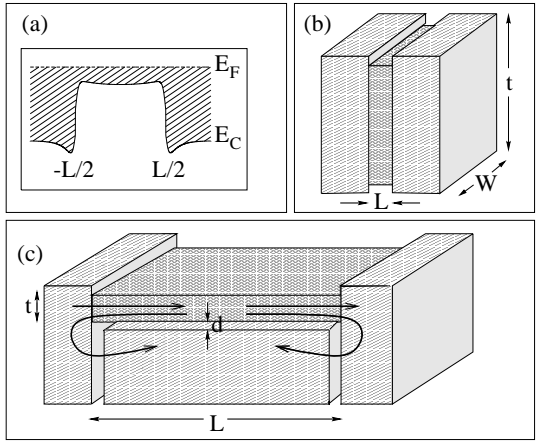,angle=0,width=80mm}}
\narrowtext
\vspace{0.5cm}
\caption{Schematic description of the geometries studied. (a) The conduction
band-edge profile in both models. (b) The sandwich model. (c) The ground-plane
model.}
\label{1model}
\end{figure}

Integration of equations (\ref{1Dcontinuity}) and (\ref{1Dpoisson})
provides a simple relation between the current and the electric field,
\[
\label{integrated}I_\omega (x)=\frac{i\omega \varepsilon 
(x)}{4\pi }E_\omega
(x)+I_{\omega}^e.
\]
The integration constant $I_\omega^e$ has the physical sense
of the current induced deep inside the electrodes (where
$E_\omega =0$). It can be found from the condition that the current
fluctuations 
do not affect the voltage $V_\omega $ applied to the structure: 
\[
\label{fixedvoltage}\int_{-\infty }^\infty dx\,\frac{4\pi }{i\omega \epsilon
(x)}\left[ I_\omega ^e-I_\omega (x)\right] =V_\omega . 
\]

Inserting equations (\ref{1Dcontinuity}) and (\ref{integrated}) in
\eq{1Dgeneralj} we get the basic equation of this model,
\[
\label{diffeq}\frac{d^2I_\omega (x)}{dx^2}-\kappa ^2(x,\omega )I_\omega (x)= 
\frac{i\omega }{D(x)}I_\omega ^s(x)-\frac 1{\lambda ^2(x)}I_\omega ^e, 
\]
with 
\[
\label{kappa}\kappa ^2(x,\omega )=\frac 1{\lambda ^2(x)}-\frac{i\omega }{D(x) 
}. 
\]
This equation is valid for both the conductor itself and the electrodes.

\subsection{The ground-plane model}
In the second (``ground plane'') model we consider a long and
thin conductor close and parallel to a well-conducting ground plane, $L\gg
t,d$ where $t$ is the thickness of the conductor, and $d$ its distance from the
ground-plane -- see Fig.~1(c). The width of the conductor $W$ (\ie, its
second dimension, parallel to the ground plane) can be arbitrary.
As we will show below, this geometry is more promising for experimental
observation of some of the effects studied in this work.

In the same way as for the previous model, equations
(\ref{continuity}) and (\ref{generalj}) can be replaced with their 1D
versions, \ie, equations (\ref{1Dcontinuity}) and (\ref{1Dgeneralj}),
respectively.  The Poisson equation, however, leads to a different 
one-dimensional equation since in this model the gradient of the field in the
$x$-direction is much smaller than the transverse gradients. In this
case, the linearity of the Poisson equation leads to a linear 
dependence of the potential on the local charge, both integrated over the
conductor's cross-section:
\[	\llabel{capacitance}
\Phi_\omega(x) = \frac{A q_\omega(x)} {C_0},
\]
where $A$ is the cross-sectional area and $C_0$ is the specific
capacitance (per unit length).  

For
a homogeneous cylindrical conductor of radius $R=t/2$
and distance $h=t/2+d$ between its center and the ground plane $C_0$
assumes the following limiting values: if the
conductor is thick ($t\gg \lambda $)\cite{Smythe 68} 
\[
C_0=\frac{\epsilon _m}{2\cosh ^{-1}\left( h/R \right) }\,\,\,\,\,\,\,\,(R
\gg \lambda), 
\]
with $\epsilon _m$ the dielectric constant of the medium between
the conductor and the ground plane. In the opposite limit, the
cylinder is uniformly charged. Taking for
simplicity $\epsilon =\epsilon _m$, one finds
\bea \nonumber
C_0 & = & \frac{2\epsilon }{1+4\ln (h/R)+4I(h/R)} \\
& \simeq & \frac{2\epsilon
}{1+4\ln 
\left( 2h/R\right) },  \spc \spc \spc (R\ll \lambda ), 
\eea
with 
\[
I(b)\equiv \int_0^1dx\,\ln \left( 1+\sqrt{1-\frac x{4b^2}}\right) . 
\]
The dependences of $C_0$ on the ratio $R/h$ are shown in Fig.~2.
\begin{figure}[tb]
\centerline{\hspace{-40pt}
\psfig{figure=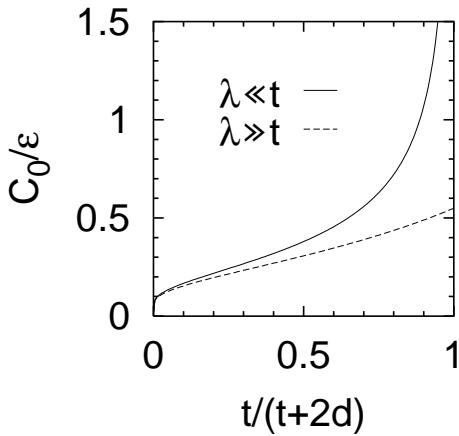,angle=-90,width=70mm}}  
\narrowtext
\vspace{0cm}
\caption{The dependence of $C_0$ on $R/h = t/(t+2d)$ for
an homogeneous cylindrical conductor close to a ground plane.}
\label{2capacitance}
\end{figure}

If the conductor is a uniform strip of width $W \gg d,
t$, then
\begin{mathletters}
\begin{eqnarray} 
C_0 & = & \frac{W}{4 \pi d}, \spc (t \gg \lambda),  \label{C0stripa} \\
	\llabel{C0stripb}
C_0 & = & \frac{W}{4 \pi (d + t/3)}, \spc (t \ll \lambda).
\end{eqnarray}
\end{mathletters}
In the quantum limit, with only the first quantum level populated,
$C_0$ may still be presented with \eq{C0stripb},
though the thickness $t$ must be replaced by an effective thickness
$t_{\rm eff}$. For a square well potential with infinite barriers
the effective thickness is
\[
t_{\rm eff} = t \left( 1 + \frac 3 {4\pi^2} \right) \simeq 1.07t.
\]

Combining Eq.~(\ref{1Dgeneralj}) with Eq.~(\ref{capacitance})
provides a diffusion-like relation between  the current and the linear
charge density, 
\[
\label{currgrpl}I_\omega (x)=- D'\frac{dq_\omega (x)}{dx}+I^s(x), 
\]
with 
\[
 D' = D + \frac{\sigma A}{C_0}.
\]
With the help of Eq.~(\ref
{1Dcontinuity}) we once again get a simple equation for the current 
\[
\label{diffeqgrpl}\frac{d^2I_\omega (x)}{dx^2}-\bar \kappa ^2(\omega
)I_\omega (x)=-\bar \kappa ^2(x,\omega )I_\omega ^s(x), 
\]
with 
\[
\label{kappabar}\bar \kappa ^2(\omega )=-\frac{i\omega }{ D'}. 
\]
Thus, as seen by comparing equations (\ref{diffeq}, \ref{kappa}) with
equations (\ref{diffeqgrpl}, \ref{kappabar}), static screening effectively
vanishes in this model, and only dynamic screening is important.

We now analyze the conditions under which \eq{capacitance} is
valid. Expanding \eq{diffeqgrpl} in spatial harmonics
we see that harmonics with wavenumber $k_x \gg
|\bar \kappa| = \sqrt{\omega/ D'}$ contribute negligibly to the current
fluctuations $I_\omega$. Thus, at frequencies $\omega
\lesssim  \bartau^{-1} \equiv D' / L^2
\ll  D'/d^2, D'/t^2$, we can consider only the wavenumbers
$k_x$ 
which are much smaller than $d^{-1}$, $t^{-1}$. For these
harmonics, the transversal gradients of the electric field dominate in
the Poisson equation, justifying Eq.~(\ref{capacitance}) at that
frequency range.

Equation (\ref{capacitance}) is not valid at distances comparable to
$d$ from the 
interface with the electrodes. Eq.~(\ref{diffeqgrpl}) should therefore be
solved only inside the conductor, with boundary conditions at $\pm 
\frac{L^{\prime }}2=\pm (\frac L2-\delta )$, where $\delta $ is now some
distance for which $d\ll \delta \ll L,1/|\bar \kappa |$. To find these
boundary conditions we write down \eq{1Dgeneralj} (which is valid even
near
the interface) in the form
\[
\label{currtrans}I_\omega (x)=-\sigma \frac d{dx}\left[ \Phi _\omega
(x)+4\pi \lambda ^2\ q_\omega (x)\right].
\]
[The sources $I^s(x)$ in the small interface
region may be neglected.] The values of $q_\omega$ and $\Phi_\omega$ at
the actual 
interfaces 
$x= \pm L/2$ are found from  the boundary conditions
Eqs.~(\ref{bcrho}, \ref{bcI}).
First note that while $q_\omega $ can be arbitrarily large on the
electrode side,  
$\lambda _eq_\omega $ should remain finite, since it is equal to the
total interface 
charge in the electrode. In the case when the
screening length in the electrodes $\lambda_e$ is much shorter than
$\lambda$, Eq.~(\ref{bcrho}) therefore gives (on the conductor side of
the interface)
\[
q_\omega \left( \pm \frac L2\right) =0.
\]
Since the
voltage drop in the electrode vanishes, the constraint of fixed
voltage [{\it i.e.}, the 
equivalent of Eq.~(\ref{fixedvoltage})] in this model becomes
\[
\Phi _\omega \left( \pm \frac L2\right) =\mp \frac{V_\omega A}2. 
\]
Integration of Eq.~(\ref{currtrans}) from $\pm L/2$ to $\pm L'/2$, and
use of Eq.~(\ref{capacitance}) at $\pm L'/2$, now yields the
required boundary conditions, 
\[
\label{bcgrpl}q_\omega \left( \pm \frac{L^{\prime }}2\right) =\mp \frac{ 
V_\omega A \sigma}{2D'}. 
\]

In this ground-plane model, finite charge densities in the conductor create
image charges on the ground plane. Thus, at any finite frequency
$\omega$ some parts of the interface currents are responsible for
periodic re-charging of the 
ground plane 
[see Fig.~1(c)]. Therefore, the current $I_\omega ^e$ measured in the
electrodes at a distance far from conductor-electrode interface may be
different from the current which flows through this interface (note
also that the two interface currents are not necessarily equal). 
In an experimental scheme symmetric with respect to the
conductor, the current $ 
I_\omega ^e$ flowing into the external circuit is the symmetric component
of the two currents:  
\[
\label{symmcurrent}I_\omega ^e=\frac 12\left[ I_\omega \left( -\frac L2 
\right) +I_\omega \left( \frac L2\right) \right].  
\]
While the currents $I_\omega(\pm L/2)$ in this expression are the
currents at the electrode side of the interfaces, due to 
Eq.~(\ref{bcI}) they are equal to $I_\omega(\pm L'/2)$. Of course, if
the leads connecting the sample to the measurement instrument have some
mutual capacitance, the current in the instrument will be less than
$I^e(\omega)$ given by \eq{symmcurrent}, but this loss factor may be
taken into account by the standard circuit theory methods.

\section{The response functions}

From now on we will consider the most natural case of well-conducting
electrodes of size much larger than $\lambda _e$, and resistance
negligible in comparison with the resistance $ R=L/\sigma A$ of the
conductor. We first show that the total noise produced in the
electrodes is negligible compared to that originating in the
conductor. For equilibrium noise, this is a direct consequence of the
fluctuation-dissipation theorem. The same is true for shot noise,
because of the 
fact that the electron distribution function in the electrodes is
almost equilibrium.
The electrode-conductor interface is also
not an appreciable source of noise since, in the diffusion
approximation studied here, the electron distribution at the
conductor-side of the interface
is the same equilibrium Fermi distribution as in the electrode. Moreover,
even deviations from this approximation would result at the most in a
few inelastic scattering events in the electrodes, leading to
thermalization of 
hot electrons arriving from the conductor. As long as $l \ll L$, the
number of those events per transfered electron is much smaller than
the number of elastic events [$\sim (L/l)^2$] the electron experiences
in the conductor, so the thermalization process at the interface can
also be neglected as a source of noise.

In this situation, the solutions to
equations (\ref{diffeq}) and (\ref{diffeqgrpl}) can thus be presented in
the form
\[
\label{jresponse} I_\omega (x)=Y(x;\omega) V_\omega + \frac 1L\int_{-\frac L2 
}^{\frac L2}K(x,x^{\prime };\omega)I_\omega ^s(x^{\prime })\,dx^{\prime }. 
\]
Eq.~(\ref{jresponse}) shows that the current at frequency $\omega$ at any
point $x$ is composed of two components. The first is the response to the
applied voltage across the conductor, and the second is the response to the
random Langevin current sources inside the conductor.

The response functions $Y(x;\omega)$ and $K(x,x^{\prime};\omega)$ are found
by solving the equations for the current with $I_\omega^s(x) = 0$ and $V_\omega=0
$, respectively. They can be presented in a compact form by defining the
following auxiliary functions, with $\eta = \kappa\lambda^2/\lambda_e$, $u =
\kappa L/2$, ${\rm sinch}(u) = \sinh(u)/u$, $\Omega = \omega/4\pi\sigma$,
and $\chi = |x| - L/2$: 
\end{multicols}
\widetext
\[
{\cal D}(\omega) = \frac{1}{\cosh(u) + \eta \sinh(u)},
\]
\[
{\cal E}_\pm(x, \omega) = \left\{ 
\begin{array}{ll}
\frac{u}{1 - i\Omega} \left[ \sinh(u \pm \kappa x) + \eta \cosh(u \pm \kappa
x) \right] & (|x| < 
\frac{L}{2}), \\ \frac{u}{1 - i\Omega} \eta \exp(-\kappa_e \chi) & (|x| > 
\frac{L}{2}), 
\end{array}
\right. 
\]
\[
\label{FF} {\cal F}(x, \omega) = \left\{ 
\begin{array}{ll}
\frac{1 - i\Omega {\cal D}(\omega) \cosh(\kappa x)}{\left[ 1 - i\Omega
\right] \left[ 1 - i\Omega {\cal D}(\omega) {\rm sinch}(u) \right]} & (|x| < 
\frac{L}{2}), \\ \frac{1 - i\Omega \left[ 1 - \eta {\cal D}(\omega) \sinh(u)
\exp(-\kappa_e \chi) \right]}{\left[1 - i\Omega \right] \left[ 1 - i\Omega 
{\cal D}(\omega) {\rm sinch}(u) \right]} & (|x| > \frac{L}{2}),
\end{array}
\right. 
\]
\[
{\cal G}_\pm(x, \omega) = \frac{i\Omega {\cal E}_\pm(x, \omega)}{\sinh(u) +
\eta \cosh(u)}, 
\]
\[
{\cal H}_\pm(x, \omega) = i\Omega {\cal D}(\omega) \left[ {\cal F}(x,
\omega) + {\cal E}_\pm(x, \omega) \right]. 
\]

\begin{multicols}{2}
For the sandwich model, the response functions are 
\bea 
\label{K} \nonumber K(x,x^{\prime };\omega ) & = & {\cal F}(x,\omega
)\pm {\cal G}_{\pm 
}(x,\omega )\sinh (\kappa x^{\prime }) \\ 
& & -{\cal H}_{\pm }(x,\omega )\cosh
(\kappa x^{\prime }) 
\eea
and 
\[
\label{Ysandwich}Y(x;\omega )=\frac{1-i\Omega }R{\cal F}(x,\omega ),
\]
where, in Eq.~(\ref{K}), the upper sign should be used for $x^{\prime }>x$
and the lower sign for $x^{\prime }<x$. 

In the ground-plane model, the functions $K$ and $Y$ inside the
conductor are found to be the same
 as for the sandwich model (but with $D \rightarrow D'$) in the limit
of vanishing 
conductivity of the conductor, {\it i.e.}, in the formal limit $\eta ^2\sim
\Omega \rightarrow \infty $ (which also implies $\kappa = \bar \kappa $):
\[
\label{Kgrpl}K(x,x^{\prime };\omega )=\cosh (\bar u\pm \bar \kappa x)\frac{ 
\cosh (\bar u\mp \bar \kappa x^{\prime })}{{\rm sinch}(2\bar u)} 
\]
with $\bar u=\bar \kappa L/2$ and with the upper sign used for $x^{\prime }>x
$ and the lower for $x^{\prime }<x$. The response in the electrodes $ 
K^e(x^{\prime };\omega )$ to a fluctuation at $x^{\prime }$ is obtained with
the help of Eq.~(\ref{symmcurrent}), 
\[
\label{Kegrpl}K^e(x^{\prime };\omega )=\frac{\cosh (\bar \kappa x^{\prime }) 
}{{\rm sinch}(\bar u)}. 
\]

The response to voltage $Y(x, \omega)$ in this model is identical to
$K^e(x; \omega)/R$,
\begin{figure}[tb]
\vspace{1.6cm}
\centerline{\hspace{0pt}
\psfig{figure=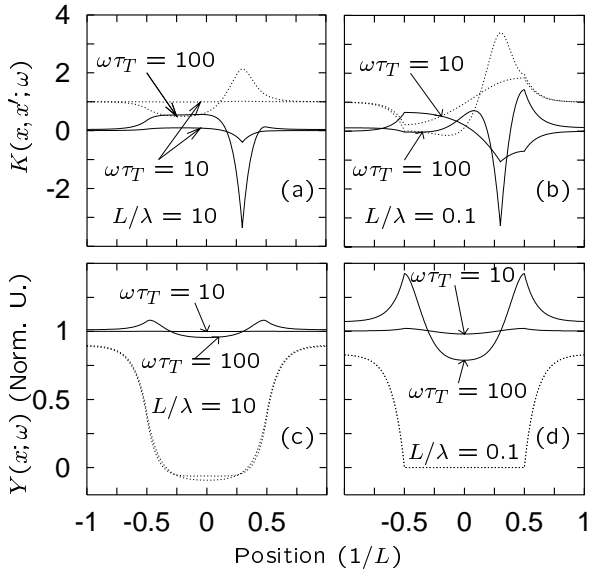,angle=-90,width=40mm}}  
\narrowtext
\vspace{0.8cm}
\caption{Position dependence of the real (solid lines) and imaginary (dashed
lines) parts of the response functions for the sandwich model at high
frequencies, and for two values of $L/\lambda$. Re$[Y(x;\omega)]$ is
in units of $1/R$. Im$[Y(x;\omega)]$ is in units of
$-\Omega/R$. $x'=0.3$ here. }
\label{3response}
\end{figure}
\[
\label{Ygrpl} Y(x;\omega) = \frac{1}{R} \frac{\cosh(\bar\kappa x)}{{\rm sinch 
}(\bar u)} . 
\]

At low frequencies, and in both models, the response functions tend to
constant values: 
\[
\label{lowfreqKY}K(x,x^{\prime };0)=1, \,\,\,\,\,\,\,\,Y(x;0)=\frac 1R. 
\]
Fig.~3 shows the response functions for the sandwich model at
intermediate and high frequencies.  At $\omega {{\tau }_T}\gg 1$
the responses are exponentially close to the source, {\it i.e.}, to
$x^{\prime }$ in the case of $K(x,x^{\prime };\omega )$ and to $\pm
L/2$ in the case of $Y(x;\omega )$. At any frequency and position, and
for any 
value of $L/\lambda$, 
\[ \llabel{intK}
\frac 1L\int_{-\frac L2}^{\frac L2}K(x,x^{\prime };\omega )\,dx^{\prime }=1. 
\]
This general result is a manifestation of the constraint $V_\omega =
0$, as can be seen by assuming a uniform current source in
\eq{jresponse}. Then, the only solution of the problem
which maintains the constraint of fixed voltage is a uniform current
everywhere, $I_\omega(x) = I_\omega^s$, leading immediately to \eq{intK}.

The relation between the spectral density of the current noise $S_I(x,\omega
)$ and the response function $K(x,x^{\prime };\omega )$ is made clear 
through the identity
\[
\label{defnoise}S_I(x,\omega )\delta (\omega )=2\langle I_\omega (x)I_\omega
^{*}(x)\rangle . 
\]
With Eq.~(\ref{jresponse}), and using the condition of fixed voltage and
the locality of the current correlator [Eq.~(\ref{intcorr})], this
expression becomes 
\[
\label{generalnoise}S_I(x,\omega )=\frac{2A}{L^2}\int_{-\frac L2}^{\frac L2 
}|K(x,x^{\prime };\omega )|^2{\cal S}(x^{\prime })\,dx^{\prime }. 
\]
The noise power deep inside the electrodes is given by 
\[
S_I(\omega )\equiv S_I(\infty ,\omega ). 
\]
The dynamic conductance is the response in the electrodes to external
voltage, 
\[
\label{defcond}Y(\omega )\equiv Y(\infty ,\omega ). 
\]

\section{Results: Conductance and equilibrium noise}

\subsection{Sandwich model}

For the sandwich model we have from equations (\ref{FF}), (\ref{Ysandwich}) and ( 
\ref{defcond}) 
\[
\label{sandcond}Y(\omega )=\frac{1-i\Omega }R\frac 1{1-i\Omega {\cal D} 
(\omega ){\rm sinch}(u)}. 
\]
If the screening length in the electrodes is very small compared to $L$ then 
$\eta u\gg 1$, and we get 
\[	\llabel{regularY}
Y(\omega )=\frac{1-i\Omega }R = R^{-1} - i\omega C, \spc C = \frac
{A} {4 \pi L},
\]
for any frequency $\omega$. Equation \ref{regularY} allows a simple
interpretation: $Y(\omega)$ is just the complex admittance of
the conductance $R^{-1}$ coupled in parallel to the capacitor $C$
formed by the two 
electrodes. 
\begin{figure}[tb]
\vspace{1cm}
\centerline{\hspace{40pt}
\psfig{figure=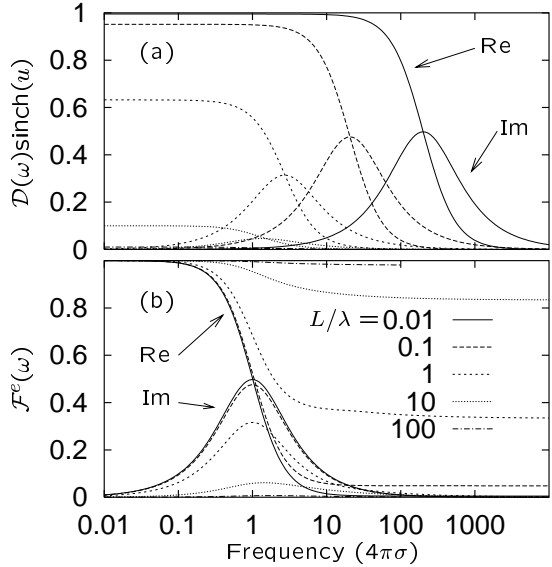,angle=-90,width=40mm}}  
\narrowtext
\vspace{1.5cm}
\caption{The real and imaginary parts of the correction to the
conductance of the sandwich model. The various curves are for
different values of 
$L/\lambda$, and for $\lambda_e = \lambda$. (a) ${\cal
D}(\omega) \sinch(u)$. The low-frequency value of this quantity is
equal to one plus the emittance of the system. (b) ${\cal F}^e(\omega)
= Y(\omega) / [(1 
- i\Omega)/R]$. }
\label{4corrcond}
\end{figure}

However, already at $L/\lambda _e=10$ the correction
\[
{\cal F}^e(\omega )= \frac 1 {1-i\Omega {\cal D}(\omega ){\rm sinch}
(u)}
\]
 to this simple result is significant. Figure 4 shows the
real and imaginary parts of this correction for different values of
$L/\lambda $, for the case $\lambda =\lambda _e$ (for example,  the
conductor and the 
electrodes are made of the same material, but the electrodes have much
fewer impurities). For $L/\lambda _e>1$ the correction term is
insensitive to an increase of $ \lambda $, so the appropriate 
curves of Fig.~4 also correspond to the case of a low-density
conductor between metal electrodes.
Fig.~4(a) shows the term ${\cal D}(\omega ){\rm sinch}(u)$ which
appears in the denominator of Eq.~(\ref{sandcond}). The low-frequency
value of this term is equal to $1-4 \pi \sigma R E(0)$, with
$\left. E(0) \equiv 
idY / d\omega \right|_{\omega = 0}$ being the
``emittance''. For a long conductor ($L/\lambda \gg 1$) 
\[ 	\llabel{emittance}
E(0) = \frac{A}{4\pi L} = \frac 1 C.
\]
 As in \eq{regularY}, the emittance in this case 
can be viewed as the sum of the intrinsic emittance of the
conductor and that of the capacitor formed by the electrodes. Then,
the total emittance (\ref{emittance}) is entirely due to the
parallel-plate capacitance, and the emittance of the ``conductor
itself'' equals 
zero, in agreement with the result of Ref.~[19]. However, as seen from
Fig.~4(a), this result does not hold for a 
relatively short conductor, which means that the simple view of an
additive emittance does not generally hold.

It is important to note that the calculations leading to \eq{sandcond}
were not dependent on the distribution function of the
electrons. Therefore, the correction to $Y(\omega )$ is due only to
the 
screening properties of the system, and does not depend on thermalization or
phase-breaking of the electrons. Equation (\ref{sandcond}) is thus
also valid 
if the inelastic scattering length is smaller than $L$. Despite its
mesoscopic nature [{\it i.e.}, the fact that  
$Y(\omega )$ assumes its ordinary value \eq{regularY} for large enough
$L$], the 
correction discussed here should not be confused with other mesoscopic
corrections to the conductivity of diffusive wires.\cite{Imry 86}

\subsection{Ground-plane model}

For the ground-plane model equations (\ref{symmcurrent}) and
(\ref{Ygrpl}) lead to the following 
expression for the conductance of the system 
\[
\label{grplcond}Y(\omega )=\frac 1 R \left(\bar\kappa \frac L 2
\right) {\rm cth} \left( \bar \kappa 
\frac L2 \right). 
\]
This expression is identical to the conductance of a macroscopic wire
of resistance $R$
coupled to the ground plane via capacitance per unit length of 
\[
\label{Ceff}C_{{\rm eff}}=\left( \frac 1{C_0}+\frac 1{C_S}\right) ^{-1}, 
\]
with 
\[	\llabel{Cs}
C_S= \frac{A}{4\pi \lambda^2}. 
\]
The boundary conditions for this model, \eq{bcgrpl}, can also be
rewritten in terms of $C_{\rm eff}$ as
\[
q_\omega \left( \pm \frac{L'}2 \right) = \frac {V_\omega}2 C_{\rm
eff}.
\]

Thus, the ground-plane model can be described by the equivalent
circuit shown in Fig.~5. The capacitance $C_S$ is due to the fact that
in thin enough wires the screening is not efficient, so the current is
determined by the gradient of the full electrochemical potential $\varphi =
\Phi + A \mu/e$, rather than by $\Phi = A q_\omega/C_0$ alone. In other
words, additional charging of the wire increases not only its
electrostatic energy ($\propto C_0^{-1}$), but also its internal
energy ($\propto C_S^{-1}$), because of the necessary rise in the
Fermi level.  In 2D conductors, Eq.~(\ref{Cs}) reduces to the
well-known result\cite{Luryi 88} for the two-dimensional electron
gas. There is also a very interesting analogy between \eq{Cs} and the
expression $L_0^{-1} = A / 4 \pi \lambda^2$\cite{Likharev 72} for the
specific kinetic 
inductance of a superconductor (in this case $\lambda$ is London's
penetration depth).
\begin{figure}[tb]
\vspace{0cm}
\centerline{\hspace{10pt}
\psfig{figure=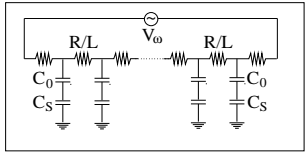,angle=0,width=40mm}}  
\narrowtext
\vspace{0.5cm}
\caption{Equivalent circuit for the geometry of the ground-plane model.}
\label{5equivcirc}
\end{figure}

When $C_s \ll C_0$ the high frequency ($\omega \bartau \gg 1$)
conductance is given by
\[
Y(\omega) = \frac 1R \frac{L}{2\sqrt{iD/\omega}}
\]
which its absolute value is just the dc conductance of a conductor of
the same conductivity 
$\sigma$, but with length equal to twice the diffusion distance in
time $1/\omega$. 
Thus, carriers injected at each of the electrodes are diffusing in and out
of the conductor, without reaching the opposite electrode, and without
affecting it by electric fields (the suppression of the longitudinal
electric fields is the only role of the ground plane in this limit).

\subsection{Equilibrium noise -- both models}

Equilibrium thermal noise is related to the conductance by the
fluctuation-dissipation theorem (we assume $T\gg \hbar \omega 
$) 
\[
S_I^{{\rm eq}}(\omega )=4T\Re \left[ Y(\omega )\right] . 
\]
However, since $Y(x,\omega)$ gives the current response to the {\it
external} voltage, the local noise $S_I^{{\rm eq}}(x,\omega )$ is not
directly related to it, and must be calculated independently. 

At zero voltage, the average distribution function in the conductor is the
Fermi-Dirac distribution given by \eq{f0}.
Using this distribution in Eq.~(\ref{SS}) gives 
\[
{\cal S} = 2\sigma T 
\]
for the correlator of the Langevin sources. The spectral density for the
equilibrium noise is thus found from Eq.~(\ref{generalnoise}),
\[
S_I^{{\rm eq}}(x,\omega ) = \frac{4T}{R L} \int_{-\frac L2}^{\frac L2}|K
(x,x^{\prime}; \omega )|^2 \,dx^{\prime };
\]
this equation in fact expresses the fluctuation-dissipation theorem
for both the local and external fluctuations.

At zero frequency Eq.(\ref{lowfreqKY}) gives
\[
S_I^{{\rm eq}}(x,0)=\frac{4T}R, 
\]
as expected.  At high frequencies ($\Omega, \omega \bartau \gg 1$), the
equilibrium noise inside the conductor is given by
\[	\llabel{eqhighfreq}
S_I^{\rm eq}(x,\omega) =  \frac {2T} {R} \sqrt{\omega \bartau /2}
\left[ f(x) + f(-x) \right] \spc (|x| < L/2)
\]
with
\[ 	
f(x) =  2e^{-\kappa_1 L}  
|\cosh(u - \kappa x) |^2 \left[ e^{2\kappa_1 x} -
e^{-\kappa_1 L} 
\right] 
\]
and with $\kappa_1 = \Re(\kappa)$, so $\kappa_1 L = \sqrt{\omega
\bartau /2}$.
$f(x) = 1/2$ throughout the conductor, except at a narrow layer of
width $1/\kappa_1$ near the edges, where it approaches its limiting
values $f(-L/2) = 0$, $f(L/2) = 2$.  The position and frequency
dependence of $S_I^{\rm eq}$ is shown in Fig. 6(a) for the case
$L/\lambda = L/\lambda_e = 10$. The general features here do not
depend on screening. 
\begin{figure}[tb]
\vspace{2.5cm}
\centerline{\hspace{-10pt}
\psfig{figure=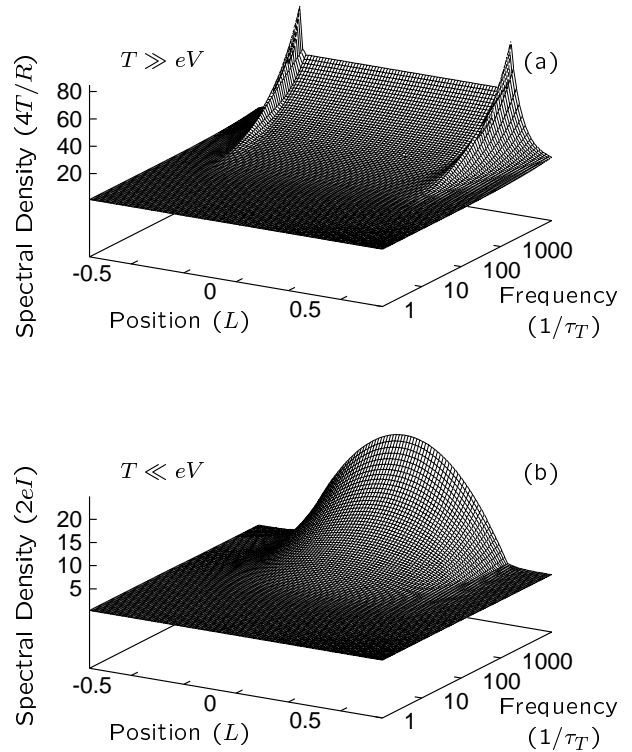,angle=180,width=55mm}}  
\narrowtext
\vspace{4.2cm}
\caption{Position and frequency dependence of the current noise intensity
inside the conductor. (a) Equilibrium noise. (b) Shot noise. Here
$L/\lambda = L/\lambda_e = 10$, but the general features of this
figure are not sensitive to the screening properties.}
\label{6spatial}
\end{figure}

The method presented here for calculating the conductance does not
depend on the form of the distribution function [other than the
diffusion approximation, Eq.~(\ref{diffapprx})]. Therefore the results
apply also to the case when strong ($l_{ee} \ll L$, with $l_{ee}$ the
electron-electron scattering length) electron-electron scattering is
present in the conductor. The equilibrium noise is also not affected
by the e-e processes since this scattering does not affect the
equilibrium Fermi-Dirac distribution, which is the input in
Eq.~(\ref{SS}).

\section{Results: Non-equilibrium noise}

In this section we will present the results on 
non-equilibrium noise for the ground-plane model; these results are also
applicable for the sandwich model in the regime $\lambda, \lambda_e
\gg L$. In the opposite limit ($\lambda, \lambda_e
\ll L$) the noise in the sandwich model is white and is equal
to the zero-frequency noise in the ground-plane model.\cite{Naveh 97}
In contrast to the case of
equilibrium noise, the shot noise is very sensitive to the strength of
electron-electron interaction in the conductor, so we analyze this
noise in the two limits of weak and strong e-e scattering.

\subsection{Weak electron-electron scattering ($L \ll l_{ee}$)}

When  $L \ll l_{ee}$ the electron distribution function
is found as a steady-state solution of
\eq{generaldiffusion}. Under the condition (\ref{smallpotential}) and for
current perpendicular to the interfaces, it
reads:\cite{Nagaev 92}
\bea	\llabel{noneqdist} \nonumber
\bar{f}_s(E,x) & = & \left( \frac 12 + \frac xL \right) f_0\left( E +
\frac 12 eV \right) \\
& & \spc + \left( \frac 12 - \frac xL \right) f_0\left( E -
\frac 12 eV \right) .
\eea
Eq.~(\ref{generalnoise}) together with equations (\ref{SS}) and
(\ref{f0}) now give a general
expression for the non-equilibrium noise,
\[	\llabel{noneqnoise}
S_I(x,\omega )  =  \frac{2}{R L}\int_{-\frac 
L2}^{\frac L2}dx^{\prime } \, |K (x,x^{\prime }; \omega)|^2 \left[
{\cal S}_+ + {\cal S}_- \frac {4x'^2}{L^2} \right]
\]
with
\[
{\cal S}_\pm(T, V) = T \pm \frac {eV}2 \cth \left( \frac {eV}{2T} \right).
\] 

Fig.~6(b) shows the spatial and frequency dependence of the
shot noise for the case $L \gg \lambda, \lambda _e$ and $T \ll eV$.
It is remarkable that inside the conductor the 
high-frequency noise is large 
even at zero temperature:
\[	\llabel{xdepshot}
S_I(x,\omega) = \frac {eI}{2} \sqrt{\frac{\omega \bartau}2} \left( 1 - \frac {4x^2} {L^2}
\right), \spc (\omega\bartau \gg 1).
\]
This rise is due to the highly non-equilibrium
distribution of carriers, \eq{noneqdist}. Specifically, at any
frequency $\omega$ the current fluctuations at position $x$ are due
only to electrons at distances $\sim \sqrt{\bar{D}/\omega}$ from
$x$. The smaller this range, the smaller is the smoothing of the
singularity in the energy distribution of the electrons in the range,
and the larger the noise.

The noise in the external electrodes is found by using the external
response function $K^e (x^{\prime }; \omega)$ in \eq{noneqnoise}.
Whenever $K^e (x^{\prime }; \omega) = 1$, \ie, at low frequencies, or
when $\lambda, \lambda_e \ll L$ in the sandwich model, \eq{noneqnoise}
reduces to 
\[	\llabel{lowfreqnoise}
S_I(\omega) = \frac{4}{R} \left[ \frac 23 T + \frac 16 eV \cth
\left( \frac {eV}{2T} \right) \right]
\]
as was found by Nagaev.\cite{Nagaev 92}
However, in the general case, the noise in the electrodes exhibits
strong frequency dependence on a frequency scale of the order of the
inverse effective Thouless time $1/{\bar \tau}_T$, which is also
affected by the screening lengths $\lambda$, $\lambda_e$.\cite{Naveh
97} Figure 7(a,b) shows the
dependence of the noise on frequency and temperature in the
ground-plane model. At $\omega \bartau \gg 1$
\[	\llabel{noneqhighfreq}
S_I(\omega) = \frac{2T}{R} \left( \sqrt{ \omega \bartau /2} - 1 \right)
+ eI \cth \left( \frac {eV}{2T} \right) + O \left( \frac 1
{\sqrt{\omega}} \right).
\]
\begin{figure}[tb]
\vspace{1cm}
\centerline{\hspace{-10pt}
\psfig{figure=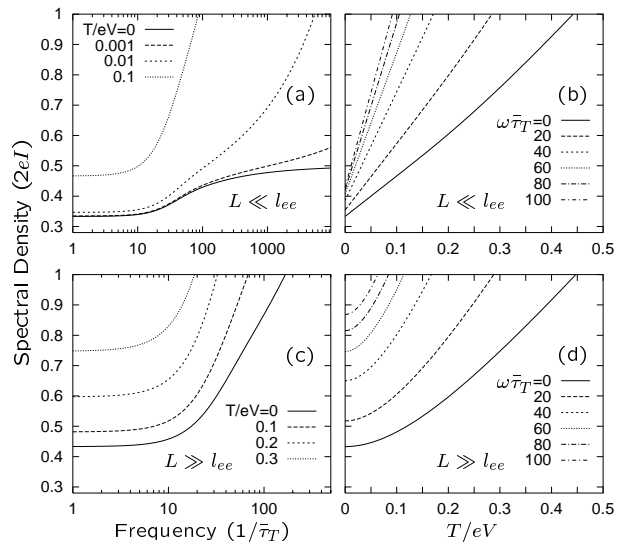,angle=-90,width=35mm}}  
\narrowtext
\vspace{1.5cm}
\caption{Frequency and temperature dependence of the spectral density of
the non-equilibrium noise in the limits of weak (a,b) and strong (c,d)
electron-electron scattering.}
\label{7tempfreq}
\end{figure}

At strictly zero temperature, \eq{noneqhighfreq} gives the
high-frequency result presented in:\cite{Naveh 97}
\[	\llabel{eI}
S_I(\omega) = eI.
\]
At finite temperatures, an additional crossover appears at
$\omega \bartau \sim (eV/T)^2/2$, above which the equilibrium noise
dominates [Fig.~7(a)]. At any frequency, \eq{noneqhighfreq} shows that
noise grows {\it linearly} with the temperature, Fig.~7(b).

The frequency dependence of the zero-temperature noise in the conductor is the
same as that of the finite-temperature noise
[compare \eq{noneqhighfreq} with \eq{xdepshot}]. Therefore, 
one can assign an effective position-dependent
temperature to the distribution function (\ref{noneqdist}):
\[ 	\llabel{Teff}
T_{\rm eff}(x) = \frac {eV}{4} \left( 1 - \frac {4x^2} {L^2}
\right).
\]
Note, however, that this distribution function does not have the  Fermi-Dirac
form.

\subsection{Strong electron-electron scattering ($L \gg l_{ee}$)}

We now consider the case when electron-electron
scattering is so strong that the scattering length $l_{ee}$ 
is much smaller than $L$ (though still larger than $l$), but is weak
enough so 
the single-particle Boltzmann equation is still valid ($l_{ee} \gg
\lambda_F$). Its solution is 
then given by the
local-equilibrium distribution
\[
f(E,x) = \frac 1 {1 + \exp \left[ \frac{E - \mu(x)}{T_e(x)} \right]}
\]
with
\[
\mu(x) = \mu - \frac{x}{L} eV
\]
and with a local electron temperature $T_e$ which satisfies
the equation\cite{Nagaev 95,Kozub 96}
\[
\frac{d^2}{dx^2} T_e^2(x) = - \frac {6 (eV)^2}{\pi^2 L^2 }
\]
under the boundary conditions $T_e = T$ at $x = \pm \frac L2$. 
The electron temperature is thus given by
\[
T_e(x) = T \sqrt{1 + \xi^2 \left[ 1 - \left( 2x/L
\right)^2 \right]}
\]
with
\[
\xi = \frac {\sqrt{3} e V}{2 \pi T}.
\]
This distribution was also found in \cite{de_Jong 96} by separating the
conductor into coupled phase-coherent conductors of length $l_{ee}$. 

With this distribution, the current correlator (\ref{SS}) becomes
\[	\llabel{See}
{\cal S}(x) = 2\sigma T_e(x)
\]
so the
non-equilibrium noise in the system is:
\[	\llabel{noiseee}
S_I(x,\omega ) = \frac{4}{R L}\int_{-\frac 
L2}^{\frac L2}dx^{\prime } \, |K_\omega (x,x^{\prime })|^2 T_e(x') .
\]
In what follows, we will concentrate on the current noise present in
the electrodes.  In all cases for which $K^e_\omega (x^{\prime }) = 1$
(\ie, $\omega \bartau \ll 1$, or $L \gg \lambda, \lambda_e$ in the sandwich model),
the result obtained in [5] is recovered:
\[	\llabel{eenoisezerofreq}
S_I(\omega) = \frac{2T}{R} \left[ 1 + \left( \xi +
\frac 1\xi \right) \arctan \left( \xi \right) \right].
\]
At high temperatures ($\xi \ll 1$) \eq{eenoisezerofreq} gives the
regular equilibrium noise $S_I(\omega) = 4T/R$. At low temperatures the
well-known result\cite{de_Jong 96,Nagaev 95,Kozub 96} 
\[	\llabel{loewfreqstrongee}
S_I(\omega) = \frac{\sqrt{3}eI}{2}
\]
is obtained.

All the above results are very close to those for $L \ll
l_{ee}$. However, the {\it high-frequency} behavior of 
non-equilibrium noise in a system with $L \gg l_{ee}$ is radically
different from that in a system with $L \ll l_{ee}$. 

Figure 7(c,d) shows the spectral density $S_I(\omega)$ as a
function of frequency and temperature for the
ground-plane model (or, equivalently, the sandwich 
model with $L \ll \lambda, \lambda_e$), with the response function
(\ref{Kegrpl}).  In the high-frequency limit, \eq{noiseee}
yields
\[ 	\llabel{highfreqee}
S_I(\omega ) = \frac TR  \omega \bartau 
\int_0^{1} e^{\kappa_1 L (y - 1)} \sqrt{1 + \xi^2 (1 - y^2)} \, dy.
\]

The characteristic parameter of this integral is
the ratio 
\[
\xi_0 = \frac {2\xi^2} {\sqrt{\omega \bartau / 2}} = \frac{3(eV)^2}
{2\pi^2 T^2 \sqrt{\omega \bartau / 2}}
\]
At small and large values of this ratio \eq{highfreqee}
gives, respectively, 
\end{multicols}
\widetext
\begin{mathletters} \llabel{limitsee}
\begin{eqnarray}
   S_I(\omega)  =  \frac{2T}{R} \left( \frac{\omega \bartau}2
\right)^{1/2} + \frac {3 R (eI)^2} {2 \pi^2 T}, & & \spc \left[ eV \ll
T (\omega \bartau)^{1/4} \right],
\llabel{limitseea} \\ 
S_I(\omega)  =  \sqrt{\frac{2 \pi^3}3} \left(
\frac{\omega \bartau}2 \right)^{3/4} \frac{T^2}{R^2 eI} +
\sqrt{\frac{3}{2\pi}} \left( \frac{\omega \bartau}2 \right)^{1/4} eI,
& & \spc \left[ eV \gg
T (\omega \bartau)^{1/4} \right].  \llabel{limitseeb}
\end{eqnarray}
\end{mathletters}
\begin{multicols}{2}

At high temperatures \eq{limitseea} is always valid. The leading term
 in $\omega \bartau$ is the same as for $L \ll l_{ee}$. 
 However, in contrast to
\eq{noneqhighfreq} the non-equilibrium correction to this 
noise has a {\it quadratic} dependence on the current $I$.

More interesting is the low-temperature limit, $\xi \gg 1$. As long as
the frequency is not very high, $\xi_0 > 1$, 
the non-equilibrium term is linear in $eI$, as usual. However, even at
$T \rightarrow 0$ the noise  grows with $\omega$ indefinitely as $(\omega
\bartau)^{1/4}$ [as opposed to the case of $l_{ee} \gg L$,
see \eq{eI}]. This dependence, and the transition to
$\omega^{1/2}$ dependence at high enough temperature or frequency are shown in
Fig.~7(c). The low-temperature behavior of the high frequency noise
can be understood by comparing the noise
correlators ${\cal S}$ in equations (\ref{noneqnoise}) and
(\ref{See}): The response function $K^e(x';\omega)$ [\eq{Kegrpl}] is
significant only for sources at $x'$ within a distance $\sqrt{D'/
\omega}$ from the interfaces with the electrodes. Therefore, the
correlator (\ref{See}), which drops more gradually near $x = \pm L/2$
than the correlator in \eq{noneqnoise}, 
produces noise in the electrodes which grows faster with frequency.

The transition to thermal noise at low temperature now occurs 
at $\omega\bartau \sim
(eV/T)^4 / 20$ (\ie, at higher frequencies than for
the case $l_{ee} \gg L$). Below this crossover \eq{limitseeb} is
valid, and the thermal term is now {\em quadratic} in $T$, Fig.~7(d).
These two results, namely, the unbound increase of the noise with
frequency at $T=0$, and the $T^2$ correction to the shot noise, are
unique to systems with strong e-e interaction, and can serve as a
clear experimental identification of such interactions.

\section{Discussion and conclusions}

We believe that our work has produced two major new results of 
general importance. First, the high-frequency noise and impedance of
small diffusive conductors is considerably affected by screening in
the conductor, the electrodes leading to the conductor, and the
surrounding media. The 'external
screening' is in fact an intrinsic part of the problem. For example,
the effects of screening on the dynamic properties are very different
[\eg, compare equations (\ref{sandcond}) and (\ref{grplcond})] for our two
models, which basically differ only in their external electromagnetic
environment. 

In particular, with due account of screening, 
the noise spectrum is {\it not} white
even at frequencies for which quantum fluctuations are negligible
and the Drude conductance is 
frequency-independent (\ie, $\omega \ll eV/\hbar, 1/\tau$).
This result is due to the fact that at
frequencies higher than the 
inverse Thouless time the current in the electrodes may be responding
only to fluctuations in the conductor which are within a distance $1/\bar
\kappa \ll L$ from the interfaces with the electrodes [see
\eq{Kegrpl}]. In these regions the distribution function is nearly
equilibrium, and therefore the noise is strongly suppressed: its
effective temperature (at $T=0$) is $eV / \bar \kappa L$,
\eq{Teff}. However, in order to satisfy the basic relation
(\ref{intK}), the external response to fluctuations in those regions
must be very large, $K^e(x';\omega) = \bar \kappa L / 2$ [see also
\eq{Kegrpl}]. Thus, each interface can be viewed as a fundamental
noise source of temperature $eV/2$,
which is just the effective temperature of the classical Schottky noise.
Since the two sources are not correlated, the total noise in the electrode
is $1/2$ of the Schottky value, \eq{eI}. [In the sandwich model, the
above description holds only for the case $L \ll \lambda,
\lambda_e$. In the opposite limit,  the strong screening allows the 
current in the electrodes to respond uniformly to all fluctuations in
the conductor, thus retaining the $1/3$ suppression factor,
\eq{lowfreqnoise}].

The reason for the peculiarity of the ground-plane model is now
apparent. Here, charge fluctuations in the conductor are screened by the
close ground plane. Therefore, high-frequency fluctuations inside the
conductor are not felt by the electrodes even if screening is strong
in the conductor and the electrodes. Thus, the frequency dispersion of
the noise is obtained in this model even if $\lambda, \lambda_e \ll
L$. 

Notice, that the result (\ref{symmcurrent}), and
therefore (\ref{eI}), is exactly valid only for the case in which the voltage
drop, and the ground plane, are symmetric with respect to the length
of the conductor. If, for instance, the ground plane is coupled much more
strongly to the right electrode, then the current through the electrodes
would be the same as the current at the left conductor-electrode
interface, and the noise value would be $2eI$. Thus, \eq{eI} is not
universal in the sense that by changing the geometry of the system any
noise value between $eI$ and $2eI$ can be obtained. 

The effect of screening on noise discussed in this work is very
different from the effect it has on classical shot noise in vacuum
diodes. In the latter case,\cite{van_der_Ziel 54} the low frequency
noise is suppressed when the space-charge in the diode (and thus the
screening) is large. This is due to the fact that, say, an upward
instantaneous fluctuation of electron  emission results in an increase
of negative space charge and hence the potential barrier near the 
emitter, so that not all excess electrons arrive at the collector. Since
the thermionic current depends exponentially on the barrier height,
this negative feedback is very effective, and the shot noise may be
considerably lower than the Schottky value.
In our case, however, this is not true. Since the Fermi level is higher
than the electrostatic  potential, electron potential fluctuations throughout
the device   hardly affect the current.

Our second major result is that 
the high-frequency noise in diffusive conductors is strongly dependent upon the
strength of the short-range electron-electron scattering. When such
scattering is strong, the non-equilibrium noise is 
not only non-white, but does not even saturate at high frequencies,
and can in fact be larger than the classical noise value $2eI$. The
quadratic dependence of the noise on temperature for this case, as
opposed to the linear dependence in the case of weak e-e scattering
(see Fig.~7), indicates that the sources of the two types of noise,
namely the thermal and shot noise, are coupled when $L \gg l_{ee}$,
but are independent (and thus additive) in the opposite
limit, $L \ll l_{ee}$. In addition to the basic significance of this
result, the different functional dependence can serve as an
experimental diagnostic tool for the determination of the ratio $\beta
= L/l_{ee}$ in a given sample. Other quantities which are sensitive to
this ratio are 
usually due to phase coherence of the electrons (and its absence at
$\beta \gg 1$) such as the corrections to the conductance due to weak
localization\cite{Altshuler 82} (as noted above, the regular, 
semiclassical conductance is not sensitive to e-e scattering). In our case,
however, the strong dependence on $\beta$ is purely semiclassical, as
it is due to the difference between the distribution functions at different
values of $\beta$.

In recent years there has been a growing experimental interest in the
dynamic properties of diffusive mesoscopic structures. While the
results presented here
are consistent with the results of all the relevant published 
experiments of which we are aware, none of those experiments explore the
regions where we predict deviations from previous theories. The ac
conductance of diffusive samples was measured at microwave frequencies
with the motivation of comparison with weak localization
theories.\cite{Vitkalov 96,Pieper 92,Sumanasekera 93} Thus, in all
those experiments the samples were very much longer than the screening
length, and a ground plane was not available. Noise measurements also
did not reach the frequency range of interest. In [33-35]
 the observation frequencies were $~ 400$ KHz
(or less). In Ref.~[36] the noise was measured at
frequencies up to 20 GHz, but the sample was made of relatively
well-conducting gold, with an inverse Thouless time of about 100
GHz. Nevertheless, we see that the experimental parameters are quite
close to those studied in this work. 

The experimental verification of the results presented in this work
should be feasible using thin conductors located very close to a
ground plane (gate).
 In many experiments this geometry is
a natural choice (\eg, when the conductor is
two-dimensional), due to its simple fabrication procedure. This
geometry also presents the 
possibility of controlling the conductor's parameters with gate voltage, cf.\
Ref.~[33]. For typical experimental
parameters,\cite{Liefrink 94} $D' \sim D \sim 10^3$ cm$^2$/s and
$L \sim 10 \mu$m, the expected crossover frequency in this geometry is
$30/2\pi \bartau \sim 5$ GHz, \ie, within the range currently
available for accurate noise measurements (cf.\ Refs.~[36,37]). 

On the other hand, the experimental observation of the noise crossover
in sandwich-type samples
can be extremely difficult, as deviations from $S_I(\omega) =
1/3$ occur in this geometry only if $L \ll \lambda,
\lambda_e$. However, observation of the screening correction to the
conductivity in this model should be possible, as this correction is
as large as $15\%$  for $L/\lambda \sim L/\lambda_e \sim 10$
[Fig.~4(b)]. With $l \sim 5 \AA$ and an electron density of $10^{15}$
cm$^{-3}$, the crossover frequency is at $2 \sigma \sim 50$ GHz.

Compared to the measurements of the noise in the electrodes,
measurements of the local noise [figures 6(a,b)] seem much more
difficult. Nevertheless, one can think of novel techniques to measure
this quantity. For instance, a measurement scheme of the local potential
spectral density, which is directly related to $S_I(x,\omega)$ by the
continuity equation, can be made possible by means of a capacitative
coupling of some point in the conductor to an external
single-electron-transistor 
probe.\cite{Schoelkopf 97a} The 
observation of this noise is then facilitated by its very large 
magnitude (Fig.~6).

\section*{acknowledgments}
Discussions with M. Buttiker, A. N. Korotkov, B. Laikhtman,
R. Landauer, Z. Ovadyahu, and R. J. Schoelkopf
are 
gratefully acknowledged. The work was supported in part by DOE's Grant
\#DE-FG02-95ER14575.

\end{multicols}

\begin{thebibliography}{99}
\bibitem{note bc} In the case when the interfaces between the conductor
and the electrodes are smooth on the scale of $l$, the diffusion
approximation applies also to the interface region, and the boundary
conditions (\ref{bcrho}, \ref{bcI}) [as well as 
their detailed form, the continuity of $f_s(\varepsilon)$ and of
$\varepsilon_\veck f_a(\varepsilon)$] can be obtained
directly by integrating \eq{generaldiffusion} across the interface region.
\end{thebibliography}
\end{document}